\documentclass{article}

\usepackage{xcolor}
\usepackage{booktabs}
\usepackage{amsmath,amssymb}
\usepackage{gensymb}
\usepackage[utf8]{inputenc}
\usepackage{lmodern}
\usepackage{newtxtext,newtxmath} 
\usepackage{tabularx}   
\usepackage{iftex}
\usepackage{textcomp}
\usepackage{graphicx}
\usepackage{forest}
\usepackage{booktabs}
\usepackage{booktabs}
\usepackage{tabularx}
\usepackage{array}
\usepackage{calc}
\usepackage{etoolbox}
\usepackage{hyperref}
\usepackage{xurl}
\usepackage[left=4cm,right=1.5cm,top=2cm,bottom=2cm]{geometry}
\usepackage{changepage} 
\usepackage{fancyvrb}
\usepackage{pmboxdraw}
\usepackage[backend=biber,style=nature,sorting=none,isbn=false,url=true,eprint=false,doi=true]{biblatex}

\hypersetup{colorlinks = true, citecolor=black, urlcolor={blue!80!black}}

\AtEveryBibitem{%
  \iffieldundef{doi}
    {}
    {\clearfield{url}}%
}

\providecommand{\tightlist}{%
  \setlength{\itemsep}{0pt}\setlength{\parskip}{0pt}}

\newcommand{\filename}[1]{\nolinkurl{#1}}

\definecolor{deepwater}{HTML}{1f78b4}
\definecolor{substrate}{HTML}{6a3d9a}
\definecolor{rock}{HTML}{59243d}
\definecolor{bedrock}{HTML}{8f4f7a}
\definecolor{boulder}{HTML}{ddd9df}
\definecolor{cobble}{HTML}{cdebde}
\definecolor{gravel}{HTML}{555654}
\definecolor{sediment}{HTML}{edcb9a}
\definecolor{sand}{HTML}{21aa9d}
\definecolor{mud}{HTML}{3e2602}
\definecolor{algae}{HTML}{417505}
\definecolor{brownalgae}{HTML}{a9680b}
\definecolor{kelp}{HTML}{dda574}
\definecolor{lamhyperborea}{HTML}{9b23ed}
\definecolor{lamdigitata}{HTML}{36489b}
\definecolor{sacchoriza}{HTML}{ca5023}
\definecolor{saccharina}{HTML}{936729}
\definecolor{alaria}{HTML}{3c2500}
\definecolor{rockweed}{HTML}{72b922}
\definecolor{rockweedhydro}{HTML}{4590e8}
\definecolor{ascophyllum}{HTML}{a7ea3e}
\definecolor{fucusvesiculosus}{HTML}{1e5faa}
\definecolor{fucusserratus}{HTML}{b34329}
\definecolor{halidrys}{HTML}{043f87}
\definecolor{rockweedgeo}{HTML}{e31a1c}
\definecolor{fucusspiralis}{HTML}{8867e9}
\definecolor{pelvetia}{HTML}{da00ac}
\definecolor{brownalgaeother}{HTML}{40321f}
\definecolor{chorda}{HTML}{0b9431}
\definecolor{desmarestia}{HTML}{417073}
\definecolor{redalgae}{HTML}{d50707}
\definecolor{corallinealgae}{HTML}{c0a38a}
\definecolor{maerl}{HTML}{f59c9c}
\definecolor{greenalgae}{HTML}{6ab019}
\definecolor{turf}{HTML}{059d4a}
\definecolor{animals}{HTML}{b15928}
\definecolor{mussels}{HTML}{e1107a}
\definecolor{mytilus}{HTML}{210399}
\definecolor{trawltrack}{HTML}{c9362a}
\definecolor{seagrass}{HTML}{417505}
\definecolor{zostera}{HTML}{417505}

\title{The MASSIMAL Dataset: UAV Hyperspectral Images and Ground Truth Observations for Coastal Habitat Mapping in Norway}
\author{Martin Skjelvareid¹*, \and Eli Rinde² \and Kasper
Hancke² \and Katalin Blix¹ \and Galice Guillaume Hoarau³ \and Maia Røst
Kile²}
\date{}

\begin{document}

\begin{adjustwidth}{2cm}{0cm} 
{\LARGE \bfseries The MASSIMAL Dataset: UAV Hyperspectral Images and Ground Truth Observations for Coastal Habitat Mapping in Norway \par}
\vspace{0.5em}
{\large \noindent Martin Hansen Skjelvareid¹*, Eli Rinde², Kasper Hancke², Katalin Blix¹, Galice Guillaume Hoarau³, Nicolo Grometto¹, Maia Røst Kile²\par}
\vspace{0.5em}
\noindent * Corresponding author, email \texttt{martin.skjelvareid@uit.no}\\
¹ UiT The Arctic University of Norway, Hansine Hansens veg 18, 9019 Tromsø, Norway \\
² Norwegian Institute for Water Research (NIVA), Sognsveien 72, 0855 Oslo, Norway.\\
³ Nord University, Universitetsalléen 11, 8026 Bodø, Norway\\

\section*{Abstract}
The MASSIMAL dataset was collected in an effort to develop new remote sensing methods for mapping shallow water coastal habitats, such as kelp forests, seagrass meadows and maerl beds. The main component of the dataset is a set of hyperspectral images collected using a unmanned aerial vehicle (UAV). Images were collected during 45 UAV flights performed at 4 main areas along the Norwegian coast, spanning across a wide latitude range, from 59$\degree$ to 67$\degree$ north. The ground sampling distance of the hyperspectral images is 3.5 cm, and the spectral range is 400-1000 nm, sampled across 300 spectral bands. During acquisition, push-broom imagery was automatically split into images with a maximum along-track length of 2000 lines. Out of approximately 1000 hyperspectral images, 221 have been annotated, using segmentation masks and a hierarchical taxonomy. An open source image processing pipeline has been developed as part of the dataset preparation, enabling efficient generation of reflectance images and glint-corrected images based on published radiance images. The dataset also includes color images (RGB) and multispectral images collected using UAV, and georeferenced image mosaics constructed from these. Ground truth data supporting the remote sensing images include hours of underwater video, thousands of geotagged underwater images, and tens of kilometers of single-beam sonar data. The data provide a foundation for assessing habitat distribution, tracking changes, and evaluating ecosystem services, thus supporting evidence-based management in Norwegian coastal environments.

\section{Background \& Summary}\label{background-summary}

\subsection{The MASSIMAL research project}
The dataset presented here was collected as part of the MASSIMAL project (Mapping of Algae and Seagrass using Spectral Imaging and MAchine Learning), whose goal was to develop methods for shallow-water coastal habitat mapping, based on hyperspectral imaging from unmanned aerial vehicles. The project was a collaboration between UiT the Arctic University of Norway, the Norwegian Institute for Water Research, and Nord University, and was conducted in the period 2020-2024. Data collection and field work was performed in the period 2021-2023. The dataset was collected during field campaigns in four areas along the Norwegian coast, chosen to represent multiple habitats of interest, and to span large variations in environmental parameters.

The data collected during the project is publicly available via the NIRD Research Data Archive \cite{Skjelvareid2025}. This includes both the primary data of interest, i.e. hyperspectral images and annotations thereof, and a large set of additional data, including RGB and multispectral images collected with a UAV, sonar measurements collected with an unmanned surface vehicle (USV), and extensive observations and underwater images used for ground truth annotation. All UAV imagery is also available via the SeaBee Geo-Visualization Portal \cite{SeaBeeGeoPortal2026}, which enables interactive viewing of the images overlayed on a background map. 

\subsection{Scientific works based on the dataset}
So far, the main scientific works based on this dataset include a conference article at the 2023 IEEE International Geoscience and Remote Sensing Symposium \cite{Skjelvareid2023} and a master's thesis \cite{Grue2022} conducted at UiT during 2022. The former demonstrates the training and validation of a U-Net model (introduced in \cite{Ronneberger2015}) for mapping marine habitats, using 81 annotated hyperspectral images from the island of Søla, within the Vega area. The latter exploits UAV hyperspectral imaging for mapping seagrass, comparing the performance of different machine learning algorithms. In addition, several posters presented results based on the dataset, at venues including the 2022 Northern Lights Deep Learning conference \cite{Skjelvareid2022}, the 2023 International Temperate Reefs Symposium \cite{Skjelvareid2023b}, and the 2023, 2024 and 2026 Marine Geological and Biological Habitat Mapping conferences \cite{Skjelvareid2023a, Skjelvareid2024, Skjelvareid2026}. 

The remainder of this article describes the dataset in more detail, including the data collection methodology, the organization and format of the dataset files, technical validation of the data, usage notes, and additional resources related to the dataset.

\subsection{Potential applications}
The dataset is designed to support development of methods for mapping coastal habitat types using optical remote sensing. Initial progress toward this goal has been achieved through the use of machine learning for habitat classification, focusing on annotated images from one of the four surveyed locations \cite{Skjelvareid2023}, yet the dataset holds substantial potential for further use in development of habitat mapping methods. This includes training models using data from all surveyed areas, and comparing multiple machine learning methods for image segmentation. 

The inclusion of extensive underwater imagery, which was recorded and used for making annotations of hyperspectral images, allows for review of existing annotations and expansion of the dataset with new annotations. In addition, unsupervised or self-supervised machine learning approaches could be devised to leverage unannotated remote sensing images, in order to harness the full potential of the information contained in the data. 

Hyperspectral, multispectral, and RGB images could serve as training data for large-scale models capable of mapping shallow-water habitats over broader geographic regions, enabling a transition from UAV-based studies to satellite and aerial remote sensing platforms. While UAV hyperspectral imagery provides fine spectral and spatial detail, its limited coverage constrains scalability; thus, an interesting application of the data would be to simulate imagery from sensors with higher coverage but lower spectral or spatial resolution \cite{Inamdar2023}. Such sensors include UAV multispectral sensors and hyperspectral satellite instruments like EnMAP \cite{Storch2023}, PRISMA \cite{Krutz2019}, or CHIME \cite{Nieke2023}. 

The dataset’s examples of strong sun and sky glint from the water surface make it valuable for developing UAV-specific glint correction methods. This is a topic well studied for satellite and aerial imaging \cite{Kay2009}, but less so for UAV imagery, which exhibits different glint patterns due to its finer spatial resolution and flexible acquisition conditions \cite{Tiskus2023,Muslim2019}. 

Since multiple scenes have been captured using both hyperspectral and multispectral sensors, the dataset enables direct comparison of these imaging modalities for shallow-water habitat mapping, supporting evaluation of factors such as radiometric calibration, georeferencing accuracy, and sun glint effects, as seen in related studies on algal bloom monitoring and precision viticulture \cite{Olivetti2023,DiGennaro2022}. 

The included sonar data further enhances the dataset’s utility, supporting algorithm development for bathymetry estimation from UAV hyperspectral imagery \cite{Lewicka2022} and enabling studies on correlations between vegetation height or density (derived from sonar) and optical imagery. 

The dataset includes temporal records of habitat extent and ecological state from selected locations surveyed repeatedly across seasons and years, providing valuable insights into habitat dynamics over time. For locations surveyed only once, the data offer a unique snapshot of habitat extent and condition at the time of observation. Collectively, these data provide an important baseline for detecting, quantifying, and interpreting future changes in Norway's marine habitats.

Beyond research, the dataset also provides educators with a practical resource for teaching high-resolution coastal mapping and geospatial data integration in fields such as remote sensing, machine learning, geography, oceanography, and ecology.

\subsection{Related datasets}
The NASA AVIRIS project has developed three generations of hyperspectral sensors for imaging from manned airplanes \cite{Green1998, Chapman2019, Green2022}, with all data openly available \cite{NASAJPL2026}. There is no centralized archive of ground truth for AVIRIS images, but some ground truth data has been published on a per-project basis, e.g. scenes such as "Indian Pines" \cite{Baumgardner2015}. Of note in the context of shallow-water habitat mapping is also PRISM, a sensor for aerial hyperspectral imaging of the coastal ocean \cite{Mouroulis2014,Heupel2013}.  Data from this mission is available via the NASA Oceandata portal \cite{Oceandata2026}, including in-situ data.  A number of hyperspectral datasets with underwater benthic images are also of interest here \cite{Dumke2017,Dumke2018b, Rashid2020}. UAV hyperspectral datasets related to shallow waters are scarce, but some datasets related to detection of plastics on-shore and in water have been published \cite{PML2023,Cefas2023,Fronkova2024}.

\section{Methods}\label{methods}

\subsection{Areas}\label{sec:areas}

The dataset was collected during field campaigns in four areas along the Norwegian coast; in the proximity of Bodø (67.2N, 15.0E), Vega (65.7N, 11.7E), Smøla (63.4N, 7.8E), and Larvik (59.0N, 10.1E). These areas were chosen to represent multiple habitats of interest (kelp forests, maerl beds, seagrass meadows) and to span large variations in environmental and geographical parameters (e.g. wave exposure, water temperature, latitude). Multiple locations were covered within each main area. An overview of areas and locations is given in Figure \ref{fig:areas_and_locations}, with symbols indicating the data types recorded at each location. 

In Bodø, flights covered a range of intertidal and shallow subtidal habitats dominated by seagrass, rockweed, kelp, and turf algae. Data were acquired over multiple years, with generally good image quality, though occasional striping and sun-glint artifacts occurred. Ground truth data was collected through snorkeling, walking, and boat-based imaging, sometimes with a temporal offset relative to UAV imaging. Some sites were revisited to enable temporal comparison. 

The Larvik sites were dominated by seagrass and brown algae. Underwater imaging for ground truth mapping was performed via snorkeling and a unmanned surface vehicle (USV), and sonar data were also acquired. The first campaign in 2021 imaged one of the sites, albeit with some technical difficulties, and a second campaign in 2023 revisited the same site while also expanding to two other sites. The 2023 campaign was affected by high concentrations of colored dissolved organic matter (CDOM) in the water, and pronounced sun and sky glint. However, conditions were otherwise good, and coverage was extensive. 

In Smøla, imaging was focused on kelp forest and sandy patch habitats. Some datasets from the Måholmen site suffered from severe problems with the inertial measurement unit (IMU) used for the hyperspectral camera, rendering them invalid except for reference imagery, while the remaining flights were successful, but influenced by rough seas, overcast weather conditions and sky glint. Ground truth data was acquired through USV underwater imaging and sonar measurements.

The Vega dataset, collected around the island of Søla in 2022, represents the most comprehensive field campaign in the MASSIMAL project, combining UAV imagery and USV sonar data with extensive ground-truth observations and underwater imagery. UAV imaging conditions were good overall, with only minor glint issues.

\begin{figure}
\centering
\includegraphics[width=\textwidth]{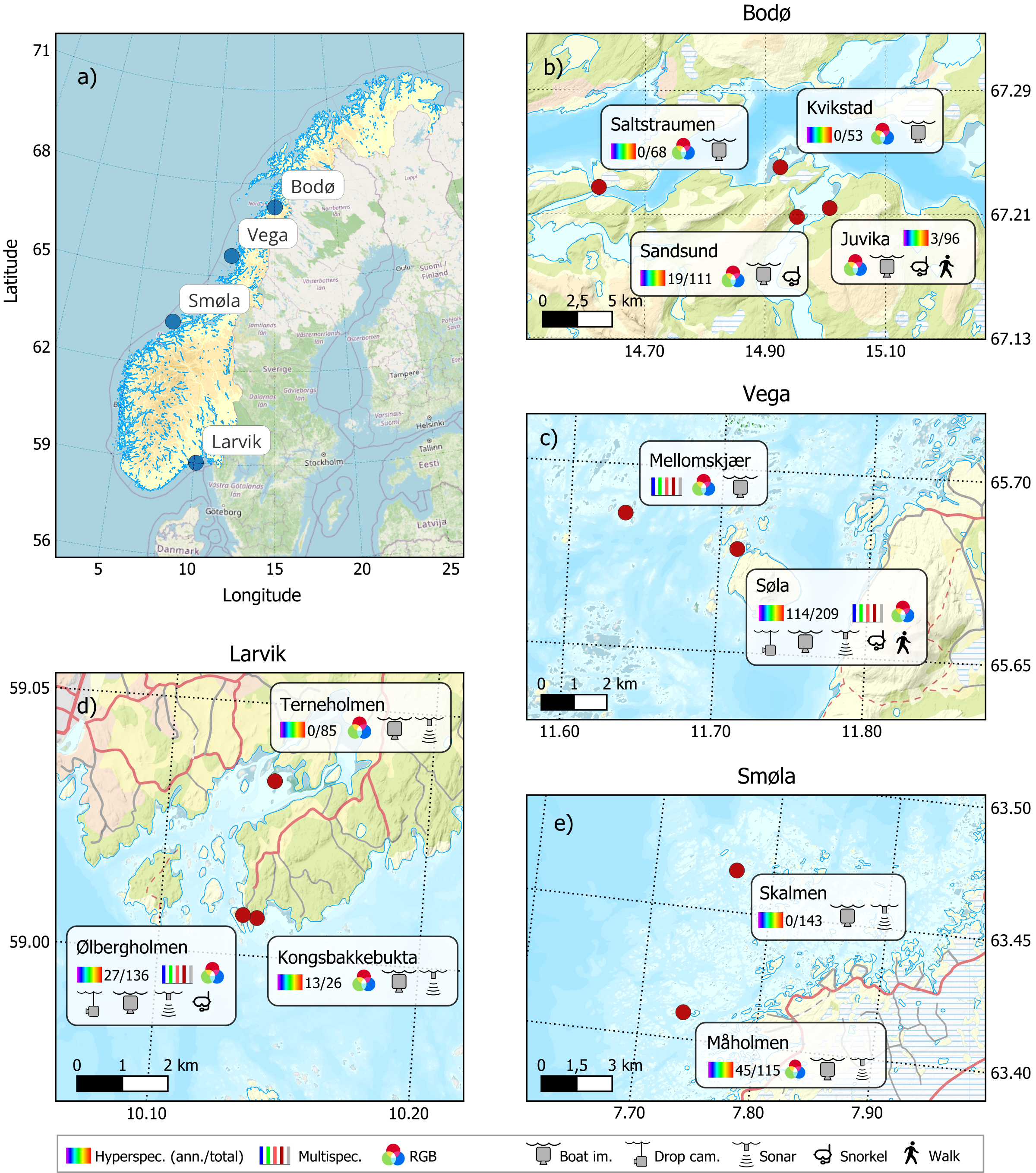}
\caption{The map in a) shows an overview of the four main areas in Norway where field work was performed, and the maps in b)-e) show specific locations within these areas. The axes of each map show latitude and longitude in degrees, but to save space, only the axes of map a) are labeled. Each location in b)-e) is annotated with the types of aerial images acquired (hyperspectral / multispectral / RGB), and ground truth sampling methods used (underwater imaging from boat or USV, drop-camera imaging, sonar, imaging while snorkeling, and imaging while walking). All maps were made available by the Norwegian Mapping Authority under the CC-BY-4.0 license \cite{CreativeCommons2013}.}
\label{fig:areas_and_locations}
\end{figure}

\subsection{Hyperspectral imaging}\label{sec:methods_hyperspectral_imaging}
Hyperspectral imaging was performed using an Airborne Remote Sensing System with a Pika-L hyperspectral camera, both manufactured by Resonon (Bozeman, Montana, USA). The hyperspectral camera and the onboard computer were mounted to a Ronin-MX 3-axis gimbal, which in turn was mounted under a Matrice 600 Pro UAV, both manufactured by DJI (Shenzhen, China). The camera was kept pointing directly downwards ("nadir") with the line of sight perpendicular to the forward movement of the drone. For yawing movements, the gimbal was set to follow the heading of the UAV, but with low angular accelerations. 

The Pika-L camera is a pushbroom sensor with 900 spatial pixels and 300 spectral channels, covering the spectral range 390-1030 nm (often rounded to 400-1000 nm). The full-width-at-half-maximum (FWHM) of the spectral channels varies between 2.24 nm and 2.94 nm across the spectral range. The camera was fitted with a lens with 8 mm focal length, giving it a 36.5 degree field of view. The camera frame rate was fixed at 100 frames (i.e., image lines) per second, and the UAV was flown at an altitude of approximately 50 m above sea level for most flights. The width of the imaged swath was approximately 33 m at the water surface, and the ground sampling distance was approximately 3.5 cm both along and across the swath. The imaging system limited the size of a hyperspectral image to 2000 lines, and automatically continued recording in a new file when this limit was reached. 

The camera gain and shutter was automatically adjusted during data collection, based on test image frames acquired by the camera, to bring the distribution of image values into a suitable part of the camera dynamic range. In some cases this resulted in suboptimal exposure settings. For example, if the UAV was above (dark) water at the time of autoexposure, and then flew over (bright) land before the exposure could be adjusted, part of the image became saturated, resulting in invalid pixels. In the post-processed images of the published dataset, partly saturated pixels are set to zero across all channels, to avoid including distorted spectra in the data.

Measurements of downwelling spectral irradiance (illumination from the sky) can be used to calculate the spectral reflectance for each pixel of a hyperspectral image. A Flame-S-VIS-NIR spectrometer with a CC-3-DA cosine collector manufactured by Ocean Insight (Orlando, Florida, USA) was used to measure downwelling spectral irradiance. The spectrometer had 2048 spectral channels covering the range of 350-1000 nm, with an optical resolution of 1.33 nm. The spectrometer was mounted directly to one of the arms of the UAV. Note that for some datasets, a technical failure (poor cable connection) resulted in downwelling irradiance not being measured.

An Ellipse 2N IMU manufactured by SBG Systems (Carrières-sur-Seine, Île-de-France, France) was connected to the onboard computer. The IMU consists of 3 accelerometers and 3 gyroscopes measuring translational and angular accelerations of the camera, a global navigation satellite system (GNSS) receiver for measuring position and velocity, a barometric altimeter for aiding altitude measurement, and a magnetometer aiding in heading measurement. The sensor data are combined in an extended Kalman filter to produce estimates of camera position (latitude, longitude, altitude) and orientation (pitch, roll, yaw). The specified accuracy for the GNSS receiver was 2.0 m CEP. Real-time or post-processing kinematic positioning (RTK/PPK) was not available for the IMU.

\subsection{RGB and multispectral imaging}\label{sec:methods_rgb_msi}
In addition to hyperspectral imaging, RGB and multispectral imaging (MSI) was performed during several of the research campaigns. These datasets were collected partly to create high-resolution base maps with higher area coverage than the hyperspectral image, and partly to compare imaging methods using the same shallow-water areas as a reference. As hyperspectral imaging is the main focus of this paper, RGB and MSI methods are only described briefly here. Interested readers are referred to detailed descriptions in \cite{Skjelvareid2026a}. 

\paragraph{RGB imaging} RGB imagery was collected using several UAV platforms, including an eBee (AgEagle, Wichita, Kansas, USA) and a Phantom 4, Mavic 2 Pro, and Matrice 210, all manufactured by DJI. The Phantom 4 and Mavic 2 Pro had integrated RGB cameras, while the others supported external sensors. The RGB images were processed using photogrammetric software such as Pix4Dmapper (Pix4D, Prilly, Switzerland) and, later in the MASSIMAL project, OpenDroneMap (ODM) \cite{PieroToffanin2024}, through the SeaBee data platform \cite{Sample2025}. Some datasets were processed with only one tool, others with both. ODM-processed datasets include both original images and derived products, while Pix4Dmapper outputs usually contain only the orthomosaic. The \texttt{report} subfolder in RGB datasets indicates which method was used. 

\paragraph{Multispectral imaging} Multispectral image data is also available in the case of two research campaigns in the Larvik and Vega areas. In Larvik (Ølbergholmen), an Altum camera (MicaSense, Seattle, Washington, USA) and a Matrice 210 UAV (DJI)  were used for imaging.  In Vega (Søla and Mellomskjær), multispectral imaging was performed by the company SpectroFly (Faxe, Denmark), using an eBee UAV (AgEagle, Wichita, Kansas, USA). The UAV was equipped with a RedEdge-M camera (MicaSense). Data was processed post-flight using AgEagle's eMotion software. The MicaSense Altum and the MicaSense RedEdge-M camera both use separate sensors for each wavelength band, including blue, green, red, red edge and NIR bands. The Altum camera includes an additional thermal band.

\subsection{Ground truth collection} \label{sec:methods_ground_truth}
During the multiple field campaigns performed in the MASSIMAL project, several methods for collecting ground truth data were developed and tested. The first campaigns (Bodø and Larvik, 2021) focused on marking relatively short transects on the seabed, and imaging these in high detail, while later campaigns (2022-2023) were more focused on collecting geotagged images across larger areas. The ground truth data was used to annotate hyperspectral images, but has also been published as part of the dataset, for two reasons: 1) for transparency regarding the annotations process, and 2) for enabling additional or modified annotations to be made. 

As hyperspectral imaging is the main focus of this paper, ground truth methods are described here with sufficient detail to understand and repeat the process, but without including all technical details. Interested readers are referred to detailed descriptions in \cite{Skjelvareid2026a}.

Many of ground truth methods are based on geotagging images by syncronizing image and position data via timestamps. A Python tool for doing this efficiently and consistently was developed as part of the project. The tool is called \texttt{vidtransgeotag} and is described in Section \ref{sec:code-availability}. 

\paragraph{USV underwater imaging and sonar measurements} An "Otter" USV (Marine Robotics, Trondheim, Norway) was used as a platform for simultaneous collection of underwater images, sonar measurements, and accurate position. Positions were measured using real-time kinetics (RTK), with cm-scale accuracy. 

A GoPro Hero 10 camera (GoPro, San Mateo, California, USA) was mounted on the USV inside a watertight casing. The camera was placed just below the water surface, close to the stern, pointing towards the sea floor. Before each recording, the internal clock of the camera was accurately set by pointing the camera at a QR code available from GoPro Labs \cite{GoProLabs2026}. This enabled accurate synchronization of the video with positions logged by the USV.

A MX Aquatic Habitat Echo Sounder (BioSonics, Seattle, Washington, USA) was integrated on the UAV. The sonar is specifically designed for acquiring information related to submerged aquatic vegetation and seafloor substrates. It has a single beam which is 8.8° wide, and a center frequency of 204.8 kHz. 

Across the whole dataset, the USV travelled a total of 52 kilometers while recording both sonar data and video. The video was processed to extract 47 000 geotagged underwater images in total, but images acquired in deeper waters contained very little information. The number of images where the seabed is at least partly visible is estimated to 40\% (\textasciitilde 19 000 images). 

The published data includes raw sonar data, which can be read and processed by the Visual Aquatic software from BioSonics. The software is freely available for registered users. All sonar data has been processed using default values in Visual Aquatic (version 1.0.0.13146), and processed data is included in the dataset as a comma-separated values (CSV) file. The main outputs of interest in the processed sonar data are:

\begin{itemize}
\tightlist
    \item Location (latitude and longitude)
    \item Time
    \item Bottom elevation (water depth, estimated from strongest echo)
    \item Plant height (estimated from time difference between first echo and bottom echo)
    \item Plant percent coverage
    \item Bottom type (based on fuzzy c-means clustering)
\end{itemize}

\paragraph{Snorkeling underwater imaging} Defined transects were marked on the seafloor before imaging. Straight transect lines were made using thin white polyester rope, or fiberglass measuring tape. Most transects had markers along them at every 5 or 10 meters. A number of different cameras were used: In datasets from Juvika and Sandsund close to Bodø, a \(\alpha\) 7R IV (Sony, Tokyo, Japan) was used. The camera was handheld, close to the water surface. In datasets from Larvik and Vega, a Hero 8 Black camera (GoPro) was used. The camera was mounted to a wooden pole, 1-4 meters long, which enabled the camera operator to lower the camera closer to the seafloor while snorkeling. This technique improved the visibility in the camera images where the water was deeper.

\paragraph{ROV underwater imaging} An miniature ROV was used during one of the research campaigns, at Søla, Vega. The ROV was a Fifish V6 Expert, manufactured by Qysea (Shenzhen, China). It includes a 4K camera and sensors for depth and temperature. The ROV was placed in the water from a boat, was piloted down to the seafloor, and driven in a straight line along the seafloor, for as far as practically possible. The start and end points of the ROV transects were measured with a handheld GPS device, resulting in a positional accuracy of a few meters.

\paragraph{Imaging from a boat} A Hero 8 Black (GoPro) camera was used for underwater imaging from a manned boat. The camera was used with a watertight casing, and was mounted to a wooden pole, which was lowered into the water along the side of the boat. The boat was driven at approximately 1 m/s (2 knots) over areas of interest, covering up to a few kilometers for a single recording.  

The position of the camera was measured using a Catalyst GNSS system (Trimble, Westminster, Colorado, USA); a lightweight GNSS receiver connected to a mobile phone. The accuracy of the system was 1-2 cm, but the GNSS receiver could not be placed directly above the camera, and the effective positional accuracy of geotagging is therefore around 1 meter.

\paragraph{Intertidal imaging} The same Hero 8 camera and Catalyst GNSS system used for imaging from a boat was also used for imaging while walking in the intertidal zone. The camera was mounted on the side of a pole and angled downwards, and the GNSS antenna was mounted on the top of the pole, directly above the camera, yielding a positional accuracy of approximately 5-10 cm.

\paragraph{Drop-camera observations} Observations of the seafloor in relatively deeper waters (approx. 5-10 m) were done in some locations, using a "drop camera", a watertight camera that is lowered down to the seafloor from a small boat. A cable attached to the camera transmitted live video back to the surface. The camera was lowered down to the seafloor, and field notes regarding depth, substrate and aquatic vegetation at a given point were made. The position of each observation was logged using a handheld GPS device.

\begin{figure}
\centering
\includegraphics[width=\textwidth]{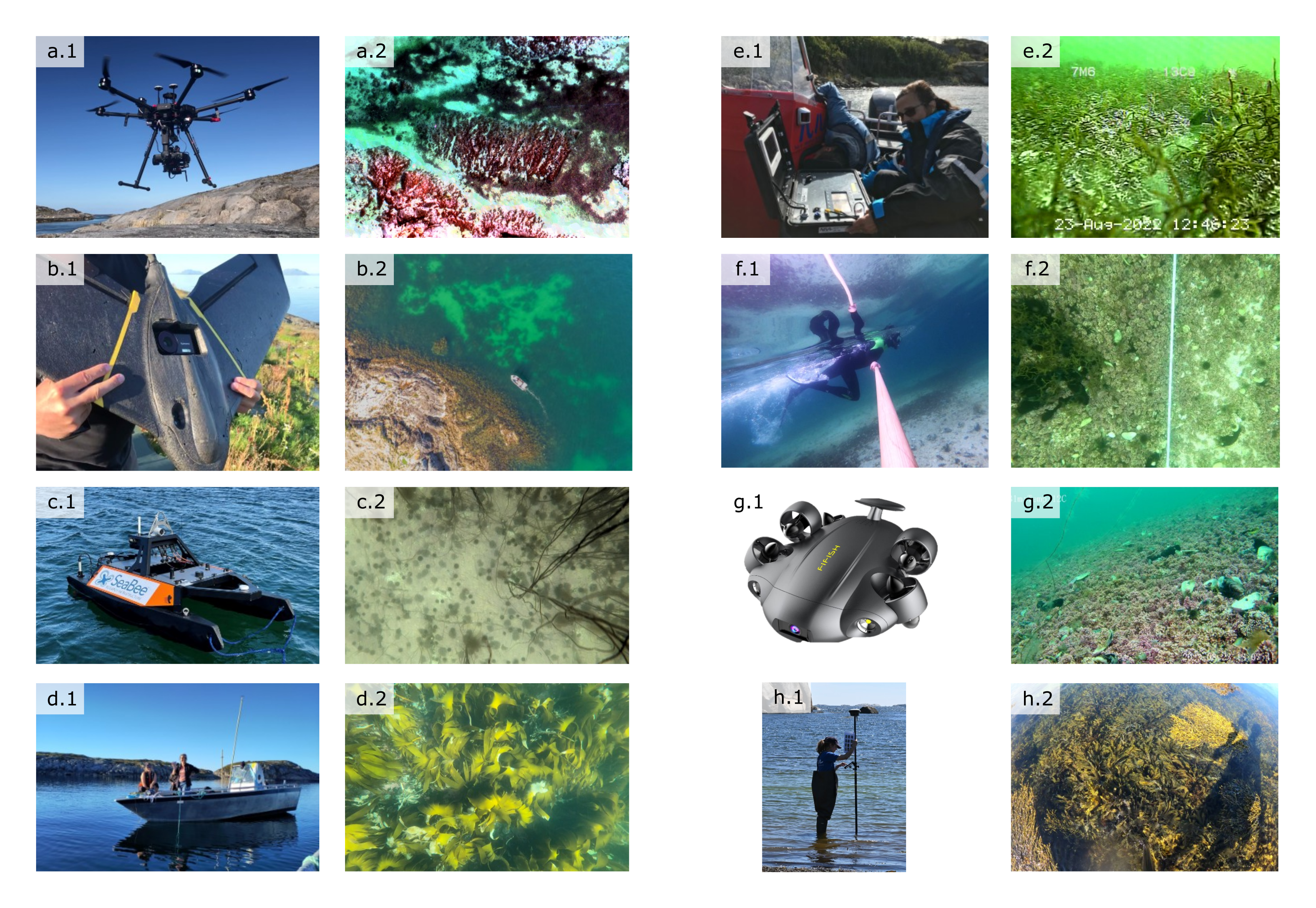}
\caption{Overview of methods used in aerial imaging and ground truth sampling methods. Image a.1 and b.1 show multicopter and fixed-wing drones, and images a.2 and b.2 show a false-color representation of a hyperspectral push-broom image and an RGB image, respectively. Image c.1 shows the "Otter" USV, and c.2 shows an example underwater image taken from the USV, showing bedrock, brown algae, and a large number of green sea urchins. Image d.1 shows a small boat used under field work, and image d.2 shows an underwater image of kelp canopy taken from the boat, using a camera mounted on a pole. Image e.1 shows the above-water monitor and electronics for the drop camera, and image e.2 shows an example image from the drop camera. Image f.1 shows a researcher snorkeling with a camera on a pole, and image f.2 is an example image of a maerl bed taken during snorkelling. Image g.1 shows the "FiFish" ROV and image g.2 shows an example ROV image, again showing a maerl bed. Image h.1 shows a person walking with a GNSS antenna in the intertidal zone, and h.2 is an example of an image of rockweed taken during a walking transect.}
\end{figure}

\subsection{Postprocessing hyperspectral images}\label{sec:postprocessing-hyperspectral-images} 
An open source data processing pipeline called \textit{MassiPipe} was developed as part of the MASSIMAL project, with the purpose of batch processing raw data into standardized data products. Details of the pipeline are given in Section \ref{sec:code-availability}.  The post-processing steps performed by MassiPipe are:
\begin{itemize}
    \item \textbf{Radiance calibration}: Calibration data supplied by Resonon (*.icp file) was used to convert hyperspectral images from raw numbers to spectral radiance, via subtraction of dark current offsets and multiplication with gain values. 
    \item \textbf{Irradiance calibration}: Calibration data supplied by Resonon (*.dcp file) was used to convert downwelling irradiance spectra from raw numbers to spectral irradiance, in the same way as for radiance images. In addition, Fraunhofer lines detected in the spectra were used to calibrate the wavelengths of each spectral channel, by fitting a polynomial to detected Fraunhofer lines with known wavelengths (similarly to the approach described in \cite{Meroni2010}). 
    \item \textbf{Irradiance resampling}: When using the irradiance spectrum for calculating reflectance images, the spectrum should be smoothed and resampled to match the spectral characteristics of the hyperspectral data \cite{Meroni2010}. Where available, irradiance spectra were smoothed with a Gaussian kernel and linearly interpolated to the wavelengths of the hyperspectral image.
    \item \textbf{IMU data interpolation and formatting}: During hyperspectral image acquisition, IMU data is logged to a text file (\texttt{*.lcf}), and timestamps for each image line is logged to a separate file (\texttt{*.times}). However, the timing of IMU data sampling and image line acquisition is not synchronized. In post-processing, IMU data is linearly interpolated to match the timestamps for each image line.
    \item \textbf{Georeferencing}: IMU data was used for georeferencing of hyperspectral images, which were mapped to geographic coordinates with a simple affine transformation (scaling and rotation). Affine transforms were generally sufficient for the MASSIMAL images because the camera was mounted on a stabilizing gimbal. Using affine transforms also has the advantage of retaining all the information of the original image.   
    \item \textbf{Mosaicing}: RGB versions of all hyperspectral images from a dataset were merged into a single mosaic, enabling viewing of the full dataset in geographic context.  
\end{itemize}
Further details about the processed data formats are given in Section \ref{sec:hyperspectral_data}. 

MassiPipe can also be used to calculate reflectance images from radiance images, if downwelling irradiance spectra are available. MassiPipe calculates reflectance as \cite{SchaepmanStrub2006}
\begin{equation}
    \rho(x,y,\lambda) = \pi \frac{L(x,y,\lambda)}{E(\lambda)}
\end{equation}
where $\rho(x,y,\lambda)$ denotes the reflectance at spatial coordinates $(x,y)$ and wavelength $\lambda$, $L(x,y,\lambda)$ denotes the hyperspectral radiance image, and $E(\lambda)$ denotes the downwelling irradiance. When calculating $\rho(x,y,\lambda)$, the single spectrum $E(\lambda)$ is broadcast across the spatial coordinates, and the resulting reflectance image $\rho(x,y,\lambda)$ requires at least as much storage space as the radiance image. Including both radiance and reflectance images in the dataset would therefore drastically increase the size of the MASSIMAL dataset. Since reflectance is easily created from the radiance and downwelling irradiance measurements, users are instead encouraged to generate reflectance images themselves. Further details on reflectance calculation is given in Section \ref{sec:usage-notes}.

\subsection{Annotation of hyperspectral images}\label{sec:annotation}
Annotation was performed by overlaying hyperspectral images with ground truth data. The QGIS software \cite{NyallDawson2026} was used to visualize multiple layers of geospatial information. These layers included: a background map layer with global coverage, a high-resolution RGB mosaic (if available), georeferenced hyperspectral images, and georeferenced ground truth data (e.g. geotagged images, transect end points, etc.).

Additional ground truth information (images, video, notes) related to the area being annotated was also reviewed in the process. With all this information available, the annotator was able to get a general impression of the area, and interpret the scene visible in the hyperspectral images. 

With geospatial information visualized in QGIS, the annotation itself was done using Hasty (CloudFactory, Berlin, Germany), an online annotation system. Annotations were made directly on RGB versions of hyperspectral images, in the form of pixel "masks" indicating class membership. Areas where the annotator had high confidence of the class were prioritized, while areas with mixed classes, transitions between classes, or low-quality ground truth were left unannotated.

Figure \ref{fig:ground_truth_and_annotations} shows an example of how multiple ground truth methods were used to cover the areas imaged from UAV, and the corresponding annotation masks. The annotation masks were georeferenced by using the same affine transform that was calculated for the hyperspectral images.   

\begin{figure}
\centering
\includegraphics[width=0.85\textwidth]{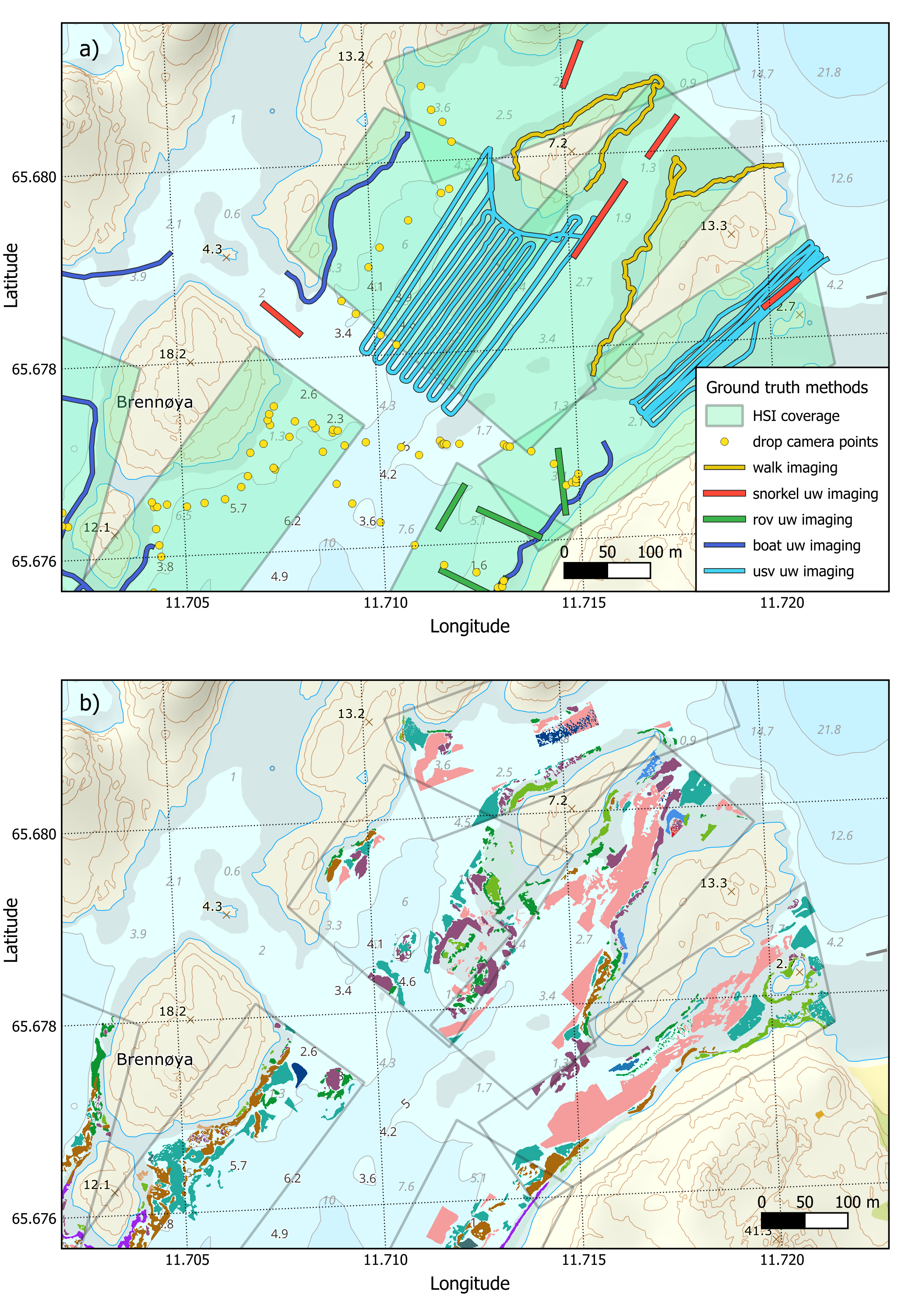}
\caption{Map a) shows a visualization of multiple methods for collecting ground truth, and map b) shows corresponding annotations made on hyperspectral images. Example from Søla, Vega. The coverage of hyperspectral images is indicated in both maps, with each polygon representing one flight (one dataset). The base map for both maps was made available by the Norwegian Mapping Authority under the CC-BY-4.0 license \cite{CreativeCommons2013}.}
\label{fig:ground_truth_and_annotations}
\end{figure}

Annotations were made using a hierarchical taxonomy shown in Figure \ref{fig:annotation-hierarchy}, inspired by the taxonomy used by the SeaBee project \cite{Kvile2024}. The hierarchical structure was used to allow the specificity of annotations to be adapted to the amount of information available: using detailed annotation classes (e.g. species classes like \textit{Fucus vesiculosus}) where accurate, high-quality information is available (e.g. high-quality underwater images or field notes), and using less detailed annotations classes (e.g. "rockweed" or "algae") where less information was available (e.g. low-quality underwater images, or UAV images only).

\begin{figure}[p!]
\begin{minipage}[t]{0.80\columnwidth}
\resizebox{0.95\linewidth}{!}{
\begin{forest}
  for tree={
    grow'=0,
    draw,
    rounded corners,
    align=left,
    edge+={->, >=stealth},
    s sep=2mm,
    l sep=4mm,
    anchor=west,
    parent anchor=east,
    child anchor=west,
    font=\sffamily\scriptsize,
    text=black,
    line width=0.5pt,
    edge={black!70}
  }
[Marine Environment
  [Deep water, fill=deepwater, text=white]
  [Algae, fill=algae, text=white
    [Brown algae, fill=brownalgae, text=white
      [Kelp, fill=kelp, text=black
        [Laminaria hyperborea, fill=lamhyperborea, text=white]
        [Laminaria digitata, fill=lamdigitata, text=white]
        [Sacchoriza polyides, fill=sacchoriza, text=white]
        [Saccharina latissima, fill=saccharina, text=white]
        [Alaria esculenta, fill=alaria, text=white]
      ]
      [Rockweed, fill=rockweed, text=white
        [Rockweed hydrolittoral, fill=rockweedhydro, text=white
          [Ascophyllum nodosum, fill=ascophyllum, text=black]
          [Fucus vesiculosus, fill=fucusvesiculosus, text=white]
          [Fucus serratus, fill=fucusserratus, text=white]
          [Halidrys siliquosa, fill=halidrys, text=white]
        ]
        [Rockweed geolittoral, fill=rockweedgeo, text=white
          [Fucus spiralis, fill=fucusspiralis, text=white]
          [Pelvetia canaliculata, fill=pelvetia, text=white]
        ]
      ]
      [Brown algae other, fill=brownalgaeother, text=white
        [Chorda filum, fill=chorda, text=white]
        [Desmarestia aculeata, fill=desmarestia, text=white]
      ]
    ]
    [Red algae, fill=redalgae, text=white
      [Coralline algae, fill=corallinealgae, text=black
        [Maerl, fill=maerl, text=black]
      ]
    ]
    [Green algae, fill=greenalgae, text=white]
    [Turf, fill=turf, text=white]
  ]
  [Seagrass, fill=seagrass, text=white
    [Zostera marina, fill=zostera, text=white]
  ]
  [Substrate, fill=substrate, text=white
    [Rock, fill=rock, text=white
      [Bedrock, fill=bedrock, text=white]
      [Boulder, fill=boulder, text=black]
      [Cobble, fill=cobble, text=black]
      [Gravel, fill=gravel, text=white]
    ]
    [Sediment, fill=sediment, text=black
      [Sand, fill=sand, text=white]
      [Mud, fill=mud, text=white]
    ]
  ]
  [Animals, fill=animals, text=white
    [Mussels, fill=mussels, text=white
      [Mytilus edilus, fill=mytilus, text=white]
    ]
  ]
  [Trawl track, fill=trawltrack, text=white]
]
\end{forest}
}
\end{minipage}%
\hspace{-15em}
\begin{minipage}[t]{0.40\columnwidth}
\vspace{-16em}
\resizebox{1.2\linewidth}{!}{
\begin{tabular}{ |p{4.8cm}||p{6.2cm}|}
 \hline
 \textbf{Grouped classes}& \textbf{Original Classes} \\
 \hline
 \textbf{Not included} & Deep water, Algae, Green algae, Turf, Animals, Mussels, Mytilus edilus, Trawl track, Human activity \\
 \hline
 \textbf{Brown Algae} & Brown algae, Kelp, Laminaria hyperborea, L. digitata, Sacchoriza polyides, S. latissima, Alaria esculenta, Rockweed, Ascophyllum nodosum, Fucus spp., Halidrys siliquosa, Chorda filum, Desmarestia aculeata \\
 \hline
 \textbf{Maerl / Red Algae} & Red algae, Coralline algae, Maerl \\
 \hline
 \textbf{Seagrass} & Seagrass, Zostera marina \\
 \hline
 \textbf{Rock} & Rock, Bedrock, Boulder, Cobble, Gravel \\
 \hline
 \textbf{Sediment} & Substrate, Sediment, Sand, Mud \\
 \hline
\end{tabular}
}
\end{minipage}
\caption{Hierarchy of marine habitats and substrates used for annotations. The table summarizes the suggested class groupings for a simplified taxonomy. Classes listed under "not included" are limited to single locations and/or a very small number of images.}
\label{fig:annotation-hierarchy}
\end{figure}

\section{Data Records}\label{sec:data-records}
The dataset has been published through the NIRD Research Data Archive \cite{Skjelvareid2025}. The data is organized as a set of ZIP archive files, organized in a directory structure with the following pattern:
\begin{Verbatim}[samepage=true]
area/
  location/
    aerial_or_surface/ 
      data_type/
        date/
          dataset/
\end{Verbatim}

Each folder may contain one or more subfolders. The \texttt{area} and \texttt{location} folders correspond to the areas and locations studied in the project, as previously described in Section \ref{sec:areas}. Data is further divided into \texttt{aerial}, meaning remote sensing from the air, and \texttt{surface}, meaning data collection performed on the sea surface or the ground. The \texttt{data\_type} indicates data type and/or data collection platform. The data types are:
\begin{itemize}
\tightlist
    \item \texttt{hsi}: Hyperspectral images collected with UAV
    \item \texttt{msi}: Multispectral images collected with UAV
    \item \texttt{rgb}: Color images (red, green, blue) collected with UAV
    \item \texttt{usv}: Underwater images and single-beam sonar data collected using an USV
    \item \texttt{boat}: Underwater images and/or written observations collected from a boat
    \item \texttt{snorkel}: Underwater images collected while snorkeling in the water surface
    \item \texttt{rov}: Underwater images collected using an ROV
    \item \texttt{walk}: Images from the intertidal zone collected while walking
\end{itemize}

The \texttt{date} folders are named using the pattern \texttt{YYYYMMDD}, with \texttt{Y}, \texttt{M} and \texttt{D} indicating the digits of year, month and day, respectively. Finally, the \texttt{dataset} folders follow the naming pattern \texttt{massimal\_\textless{}area\textgreater{}\_\textless{}location\textgreater{}\_\textless{}datetime{[}optional\ tag{]}\textgreater{}\_\textless{}data\ type\textgreater{}}, summarizing the same information as the directory tree. Note that for some datasets, the \texttt{datetime} includes minutes (\texttt{m}) and seconds (\texttt{s}), using the pattern \texttt{YYYYMMDDmmss}, while some only include the date. Including the time is useful for separating datasets from multiple UAV flights performed on the same day. Some datasets also include a "tag" (e.g. "south" or "coast2") to more easily identify a dataset. Note that annotation of hyperspectral images are placed under the \texttt{hsi} data type. Datasets with hyperspectral images are placed under the subfolder \texttt{\textless{}dataset\textgreater{}/processed}, for reasons related to publication on the SeaBee Geo-Visualization Portal.

\subsection{Hyperspectral Image Data}\label{sec:hyperspectral_data}
Table \ref{tab:hyperspectral_datasets} provides an overview of hyperspectral image data, with locations, dates, number of images, and details on ground truth collection.

\begin{table}[p]
  \centering
  \begin{tabularx}{\textwidth}{@{}
      >{\raggedright\arraybackslash}p{0.16\textwidth}
      >{\raggedright\arraybackslash}p{0.13\textwidth}
      >{\centering\arraybackslash}p{0.07\textwidth}
      >{\centering\arraybackslash}p{0.09\textwidth}
      >{\centering\arraybackslash}p{0.12\textwidth}
      >{\centering\arraybackslash}p{0.03\textwidth}
      >{\raggedright\arraybackslash}X@{}}
    \toprule
    \textbf{Location} & \textbf{Date} & \textbf{Flights} & \textbf{Images} & \textbf{Annotations} & \textbf{DI} & \textbf{Ground truth} \\
    \midrule
    \textbf{Bodø} & & & & & & \\
    Juvika & 2021-04-18 & 4 & 60 & 0 & Yes & Snorkeling \\
    & 2022-06-24 & 1 & 36 & 3 & Yes & Walking \\
    Kvikstad & 2023-10-13 & 2 & 53 & 0 & 1/2 & Boat imaging \\
    Saltstraumen & 2022-03-12 & 3 & 68 & 0 & Yes & Boat imaging \\
    Sandsund & 2021-04-18 & 1 & 19 & 0 & Yes & Snorkeling \\
    & 2021-05-20 & 2 & 29 & 0 & Yes & Snorkeling, boat imaging \\
    & 2021-06-02 & 1* & 31 & 19 & Yes & Snorkeling, boat imaging \\
    & 2023-10-13 & 1 & 32 & 0 & No & Boat imaging \\
    \addlinespace[0.5em]
    \textbf{Larvik} & & & & & & \\
    Kongsbakkebukta & 2023-08-30 & 1 & 26 & 13 & Yes & USV imaging and sonar \\
    Ølbergholmen & 2021-08-25 & 3* & 84 & 18 & No & Snorkeling, boat dropcam \\
    & 2023-08-30 & 3 & 52 & 9 & 2/3 & USV imaging and sonar \\
    Terneholmen & 2023-08-31 & 3 & 85 & 0 & No & USV imaging and sonar \\
    \addlinespace[0.5em]
    \textbf{Smøla} & & & & & & \\
    Måholmen & 2023-06-19 & NA & (159) & 0 & NA & Problems with georeferencing make ground truth unusable. Not considered a valid dataset, but quicklook images included for reference. \\
    & 2023-06-21 & 6 & 115 & 45 & Yes & USV imaging and sonar, towed camera \\
    Skalmen & 2023-06-20 & 7 & 143 & 0 & Yes & USV imaging, sonar \\
    \addlinespace[0.5em]
    \textbf{Vega} & & & & & & \\
    Søla & 2022-08-23 & 7 & 209 & 114 & Yes & Snorkeling, USV imaging, sonar, boat (dropcam \& imaging), ROV, walking \\
    \addlinespace[0.5em]
    \textbf{Total} & & 45 & 1042 & 221 & & \\
    \bottomrule
    \multicolumn{7}{@{}p{\textwidth}@{}}{\footnotesize
      Flights marked with an asterisk (*) indicate datasets with multiple versions (original and orthorectified).}
  \end{tabularx}
  \caption{Overview of all hyperspectral datasets, with areas, locations, number of flights (each flight corresponds to one dataset), the total number of images per location, the number of annotated images per location, whether downwelling irradiance (DI) was measured, and what types of ground truth were collected.}
  \label{tab:hyperspectral_datasets}
\end{table}

Each hyperspectral image dataset (each "flight") has the following structure:

\begin{Verbatim}[samepage=true]
    ├── 1a_radiance
    │   ├── rgb    
    │   │   ├── <DatasetName>_<ImageNumber>_radiance_rgb.tiff
    │   │   └── ...
    │   ├── <DatasetName>_<ImageNumber>_irradiance.spec
    │   ├── <DatasetName>_<ImageNumber>_irradiance.spec.hdr
    │   ├── <DatasetName>_<ImageNumber>_radiance.bip
    │   ├── <DatasetName>_<ImageNumber>_radiance.bip.hdr
    │   └── ...
    ├── imudata
    │   ├── <DatasetName>_<ImageNumber>_imudata.json
    │   └── ...
    ├── orthomosaic
    │   └── <DatasetName>_<ImageType>_rgb.tiff
    ├── quicklook
    │   ├── <DatasetName>_<ImageNumber>_quicklook.png
    │   └── ...
    ├── config.seabee.yaml
    ├── license.md
    └── readme.md
\end{Verbatim}

Note that raw hyperspectral data is not included in the dataset, as it would require significant amounts of storage space without adding much value to the dataset. Applications of hyperspectral data are usually based on calibrated radiance data, or further processed data, e.g. reflectance.       

The details of each type of data in the dataset is described below. The presentation follows the subfolder structure of the dataset. 

\paragraph{Radiance images and irradiance spectra} The \texttt{1a\_radiance} subfolder contains hyperspectral images representing spectral radiance, as well as downwelling irradiance spectra (where available). This data constitutes the largest and most important part of the dataset. The radiance images are saved in the ENVI format \cite{EnviHeader2026}, which splits the data into two files; a binary file with extension \texttt{*.bip}, and a header text file with extension \texttt{*.bip.hdr}. The header file contains metadata about the image, such as the size of the image and the wavelengths corresponding to each spectral band. The radiance image headers also include the field "map info", which provides georeferencing information for the image. The georeferencing format is compatible with commonly used tools such as QGIS and GDAL. 

The spectral radiance values are in units of microflicks $(10^4 \ \text{W}\cdot\text{sr}^{-1}\cdot\text{m}^{-3})$. Although not very common, this unit is used here because it scales typical values to a range well suited for encoding as 16-bit unsigned integers. To scale radiance from microflicks to the more common unit of $\text{W}\cdot\text{sr}^{-1}\cdot\text{m}^{-2}\cdot\text{nm}^{-1}$, divide values by $100\ 000$.

Irradiance spectra are saved in the same ENVI format as the hyperspectral images, but with file name extensions \texttt{*.spec} and \texttt{*.spec.hdr} for the binary and header files, respectively. Irradiance is saved with units $\text{W}\cdot\text{m}^{-2}\cdot\text{nm}^{-1}$. 

For datasets where downwelling irradiance was measured, a version of the irradiance spectrum has been written to the header of the radiance image, under the field "solar irradiance". Irradiance values are in W/(m²·µm), the default unit for solar irradiance in ENVI files. This version of the irradiance spectrum has been smoothed and resampled to match the hyperspectral camera's wavelengths (see Section \ref{sec:postprocessing-hyperspectral-images}). 

\paragraph{Radiance RGB GeoTIFFs} The folder \texttt{1a\_radiance} has a subfolder named \texttt{rgb}, which contains RGB versions of the radiance images, saved as georeferenced GeoTIFF files. These files can be useful as lightweight representations of the hyperspectral image, and are saved with the same physical units as the original radiance image (microflicks). The red, green and blue image channels are bands from the hyperspectral image corresponding to 640, 550 and 460 nm, respectively. 

\paragraph{IMU data} Data in \texttt{imudata/} is stored as JSON files, with key-value pairs for 7 different measurements. Each value is a numeric array, and the length or the arrays match the number of lines in the corresponding hyperspectral image. The \texttt{time} array represents time as a single floating-point value, describing the number of seconds passed since January 1st 1970 ("UNIX time"). The \texttt{roll}, \texttt{pitch}, and \texttt{yaw} arrays represent the orientation of the camera, measured in radians. Positive roll and pitch values correspond to pointing the camera to the right and forwards, respectively. Yaw is zero at due north, $\pi/2$ at due east. Coordinates \texttt{latitude} and \texttt{longitude} are measured in in decimal degrees, and are positive for northern and eastern hemispheres, respectively. Finally, \texttt{altitude} is measured in meters relative to the WGS-84 ellipsoid.

\paragraph{Mosaic} The \texttt{orthomosaic} folder contains a single RGB GeoTIFF file for visualization of all the images in the dataset. The mosaic is a combination of all the radiance RGB GeoTIFFs (in \texttt{1a\_radiance/rgb}), resampled to a unified spatial resolution. The mosaic has also been percentile stretched for enhanced contrast, and is saved in an 8-bit file format, and thus does not retain the physical units of the original images. 

\paragraph{Quicklook} The images in \texttt{quicklook/} images are color images saved in PNG format, and are useful for getting a quick overview of the dataset. The PNG image format allows the images to be opened by any standard image viewer. The red, green and blue (RGB) color channels of the images correspond to slices of the hyperspectral images at wavelengths 640, 550 and 460 nm, respectively. The images have also been contrast stretched, and do not retain the physical units of the hyperspectral images. 

\paragraph{Root level files} The configuration file \texttt{config.seabee.yaml} contains metadata about the dataset, and the parameters for processing the dataset using MassiPipe (see Section \ref{sec:usage-notes}). The \texttt{license.md} file contains licensing information CC BY 4.0), and the \texttt{readme.md} file contains information about the MASSIMAL research project and technical details about the hyperspectral data.

\vspace{0.5em}

\subsection{Annotation data}\label{sec:annotation_data}
Annotation datasets are named \filename{massimal_<area>_<location>_<date>_hsi_annotation}, and contain annotations for all hyperspectral datasets collected on a given date. The data is organized with the following file structure:
\begin{Verbatim}[samepage=true]
    ├── rgb_images
    │   ├── <dataset_alias>
    │   │   ├── <dataset_name>_<image_number>.png
    │   │   └── ...
    │   └── ...
    ├── segmentation_masks
    │   ├── label_classes.json
    │   ├── <dataset_alias>
    │   │   ├── <dataset_name>_<image_number>.png
    │   │   └── ...
    │   └── ...
    ├── annotations_json
    │   ├── <dataset_base_name>_coco.json
    │   └── <dataset_base_name>_hasty.json
    └── readme_hsi_annotation.md   
\end{Verbatim}

The \texttt{rgb\_images} folder contains the original images that the annotations were based on. The images are "quicklook" images, i.e. contrast stretched individually for each color channel. Some images have also been glint corrected (see Section \ref{sec:usage-notes}) to increase the visibility of the seabed. 

The \texttt{segmentation\_masks} folder contains PNG images with annotations color coded according to class. Black indicates non-annotated pixels. The folder also contains \texttt{label\_classes.json}, which describes the different annotation classes, their names and descriptions, and their color.  

For reasons related to internal file organization in the annotation tool, annotations from the same date were grouped together. The images in \texttt{rgb\_images} and \texttt{segmentation\_masks} are organized in subfolders (\texttt{<dataset\_alias>}) that correspond to hyperspectral datasets. The folder names don't correspond exactly with the dataset names, but the names of the PNG image files \textit{do} correspond to the names of the hyperspectral images. 
 
The \texttt{annotations\_json} folder contains two JSON files, exported in "Hasty" and "COCO" format, and are alternatives to using segmentation masks. The COCO format is well established and supported by multiple libraries and tools \cite{Lin2014}. The proprietary Hasty format \cite{HastyJson2026} is based on the COCO format, but enables additional metadata, which in the MASSIMAL dataset is used to add attributes to some of the annotations. Both formats define the annotation masks directly in the JSON file, i.e., segmentation masks for multiple images are contained in a single file.

\subsection{RGB and multispectral aerial images}\label{sec:additional_aerial_data}

RGB and multispectral images have been processed with photogrammetry software (either Pix4Dmapper or OpenDroneMap) to create georeferenced othomosaics. The dataset structure is the same for both modalities:

\begin{Verbatim}[samepage=true]
    ├── dem/
    ├── images/
    ├── orthophoto/
    ├── other/
    ├── report/
    └── config.seabee.yaml
\end{Verbatim}

Where available, the original images collected by the UAV are saved under \texttt{images}. The \texttt{orthophoto} folder contains RGB orthomosaics in GeoTiff format. The same mosaic has been saved in multiple file versions in conjunction with publication of the data on the SeaBee Geo-Visualization Portal \cite{Sample2025}. The \texttt{dem} folder contains digital elevation models. The photogrammetry software was not designed for calculating under-water elevation (bathymetry), but in some shallow areas the software is still able to estimate the shape of the seabed under water. The \texttt{report} folder contains a PDF report with an overview of the photogrammetry processing, and statistics describing the quality of the outputs. The \texttt{other} folder contains additional or non-standard files. The file \texttt{config.seabee.yaml} contains metadata about the dataset, including location, time, number of images, and photogrammetry processing parameters. 

\subsection{Ground truth datasets}
\label{sec:ground_truth_datasets}
As hyperspectral imaging is the main focus of the present dataset and paper, the format for ground truth data is described only briefly here. Additional details can be found in \cite{Skjelvareid2026a}.

\paragraph{USV underwater images and sonar data} The USV was used for collection of two data modalities simultaneously; underwater imaging and single-beam sonar measurements. The two modalities are published together as a single dataset with the following file structure:

\begin{Verbatim}[samepage=true]
    ├── geotagged_images/
    |   ├── images/
    |   └── <dataset_name>_geotagged_images.gpkg
    └── sonar/
        ├── raw/
        └── processed/
\end{Verbatim}

The \texttt{geotagged\_images} folder contains underwater images saved as *.jpg files, and a *.gpkg (GeoPackage) file with the positions of each image. The attribute table for each point in the *.gpkg file contains the corresponding file name for the image. The \texttt{sonar} folder contains raw sonar data saved as *.rtpx files (a proprietary file format), and processed sonar data formatted as a single CSV file. The CSV includes columns with time, position, water depth and vegetation height data.

\paragraph{Snorkeling images and video} Datasets containing images and/or video of the seafloor taken while snorkeling are organized with the following main folders. 

\begin{Verbatim}[samepage=true]
    ├── images/
    ├── video/
    └── transect_positions/
\end{Verbatim}

When these datasets were created, no standardized methodology had been established, and each dataset is therefore formatted slightly differently. Some datasets contain only images or only video. The transect\_positions folder contains a GeoPackage (*.gpkg) or CSV file (*.csv) defining the names and geospatial endpoints of the transects. The files and folders under \texttt{images} and \texttt{video} uses the same transect names.

\paragraph{Walk images} Datasets containing geotagged images collected while walking in the intertidal zone are organized in the same way as USV underwater imaging data:
\begin{Verbatim}[samepage=true]
    └── geotagged_images/
        ├── images/
        └── <dataset_name>_geotagged_images.gpkg
\end{Verbatim}

\paragraph{Imaging / observations from boat} Datasets containing geotagged images collected by underwater imaging from a boat are organized in the same way as USV underwater imaging data:
\begin{Verbatim}[samepage=true]
    └── geotagged_images/
        ├── images/
        └── <dataset_name>_geotagged_images.gpkg
\end{Verbatim}
Some data from drop-camera observations are also included in the \texttt{boat}, but these were not standardized and will not be described here. Readers are referred to \cite{Skjelvareid2026a} for details.

\paragraph{ROV video} Only a single ROV video dataset was recorded. It includes ROV video from field work at Søla, Vega, and is organized as follows:
\begin{Verbatim}[samepage=true]
    ├── video/
    │   ├── rov_transect_1
    │   └── ...
    └── massimal_vega_sola_20220823_rov_transect_endpoints.gpkg
\end{Verbatim}

Each video folder contains a video file from a single transect. The GeoPackage file contains the endpoints for every transect.

\section{Data overview}\label{sec:data-overview}
The dataset includes a large amount of hyperspectral data, and a detailed analysis is outside the scope of this data descriptor. However, we include a short overview of spectral properties here. Due to practical computer memory constraints, the overview is based on a subset of all annotated pixels in the dataset, with 2,5\% of annotated pixels sampled from each image, totaling 2.6 million spectra. The overview is based on the five groups on annotated classes (sediment, rock, maerl, seagrass, brown algae) listed in Figure \ref{fig:annotation-hierarchy}. 

Figure \ref{fig:spectral_statistics} shows percentile statistics for the reflectance spectra in the dataset. The overall statistics (upper left plot) show that reflectance values in the visible range (400-700 nm) are mostly limited to the range 1-10\%. In the near-infrared range (> 700 nm), the dynamic range of reflectance expands to around 0-50\%, which is mainly due to two effects: above-water (intertidal) brown algae reflect strongly in the near infrared, increasing the upper bound of the values, and near-infrared light from below-water habitats is effectively absorbed by the water, decreasing the lower bound of values. Within the visible range, spectra follow a general trend of values increasing from 400 nm to a peak around 580 nm, and decreasing towards 700 nm. This is probably mainly due to scattering and absorption effects in the water, which typically makes Norwegian waters most transparent for green light (520–570 nm). The spectra are quite similar and overlapping for the five grouped classes, indicating that separation of classes is not straightforward, but some classes also have distinct spectral features. Also, a small peak is seen around 700 nm in the median spectrum for many of the classes. This is caused by the "red edge" where plants and algae transition to high NIR reflectance. While plants on land have a high-reflectance plateau in the NIR range, absorbance in water causes the plateau to taper off quickly, resulting in a small local peak.  
\begin{figure}[!t]
\centering
\includegraphics[width=\textwidth]{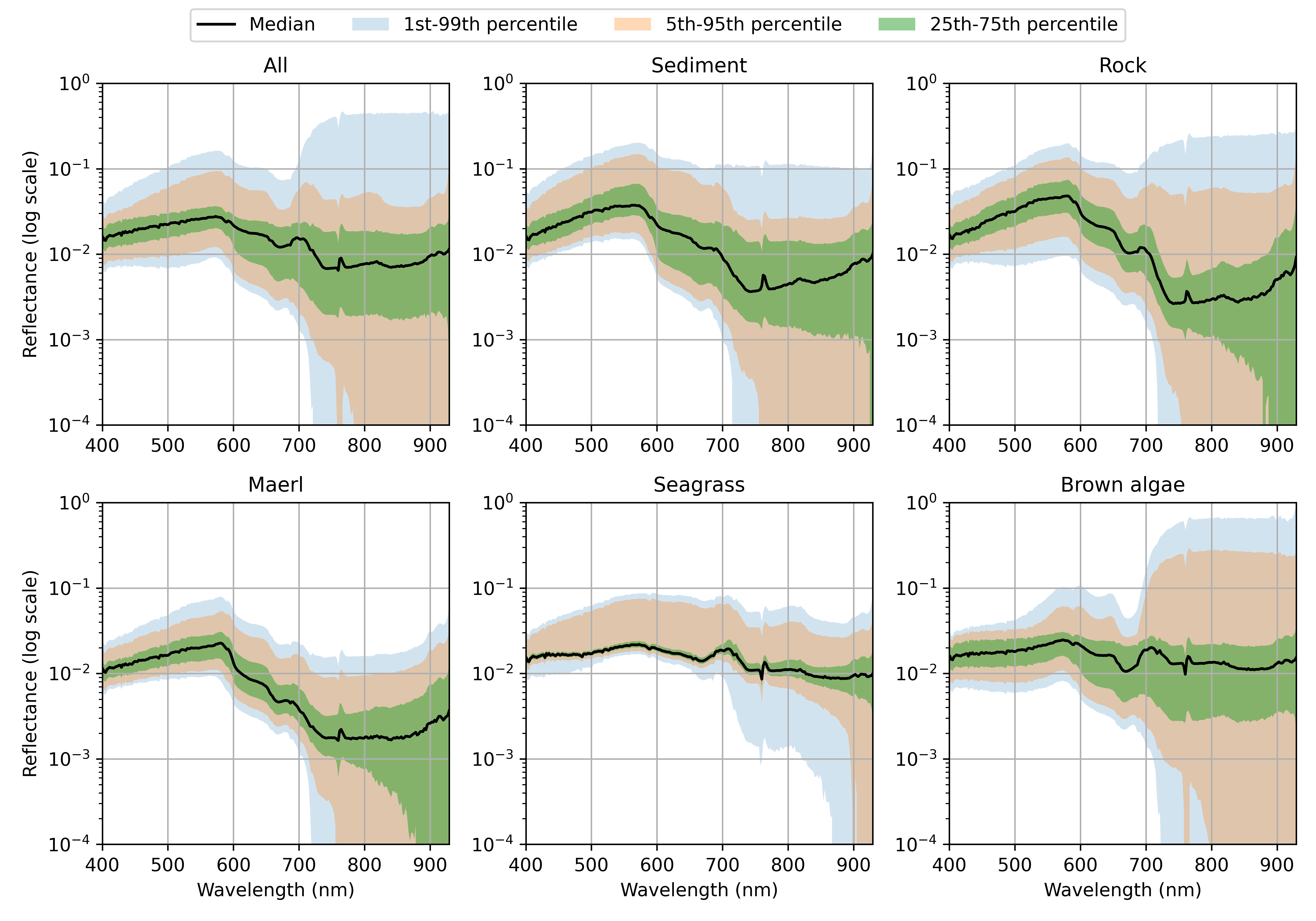}
\caption{Spectral statistics for annotated regions of hyperspectral reflectance images. The upper left plot shows statistics for the the dataset as a whole, and the remaining plots shows statistics for five grouped classes.}
\label{fig:spectral_statistics}
\end{figure}

Figure \ref{fig:umap} is a visualization of the distribution of spectra in high-dimensional space. Each spectrum has been reduced to two dimensions via principal component analysis (PCA) and subsequent uniform manifold approximation and projection (UMAP) \cite{McInnes2018}. Before PCA, spectra were smoothed using a Savitzsky-Golay filter, and normalized as z-scores for each wavelength. Nine PCA components were used, representing 99.7 percent of the dataset variance. A 2D UMAP model was fitted to 0.025 \% of all the annotated spectra, as a smaller subset enabled both faster training and a less cluttered visualization. The full 2,5 \% subset was then transformed to UMAP coordinates using the trained model. The UMAP method attempts to visualize data points such that points that are close in high-dimensional space remain close in the 2D visualization. UMAP transformed points were visualized as an aggregated scatter plot using DataShader \cite{JamesBednar2026}. In Figure \ref{fig:umap}, pixels were colored based on grouped class membership, and histogram equalization was used for contrast enhancement. 

The spectra generally don't separate into clearly defined clusters. Some classes are distributed across large parts of the feature space, and there is clear overlap between some classes. In addition to the natural variation in reflectance for each class (which includes multiple subclasses), the large inter-class variations are probably caused by effects such as varying water depth and water surface reflections. Seagrass is the class that is best separated from the others, with most samples located in a small cluster. Note, however, that the dataset contains relatively few annotations of seagrass, with the majority being from a single location (Ølbergholmen, Larvik). Sediment and rock are the two classes with the highest degree of overlap, which is to be expected, as the sediment is often derived directly from the parent rock. Some spectra from brown algae and maerl are clearly separated from others, and these are probably from shallow areas with rich spectral detail, while others are more overlapping, probably corresponding to deeper areas with weaker signals and an narrower spectral range. Note that figure \ref{fig:umap} represents spectral information only, and that adding textural (spatial) information can help increase the separability of classes and improve habitat mapping accuracy.

\begin{figure}[!t]
\centering
\includegraphics[width=0.80\textwidth]{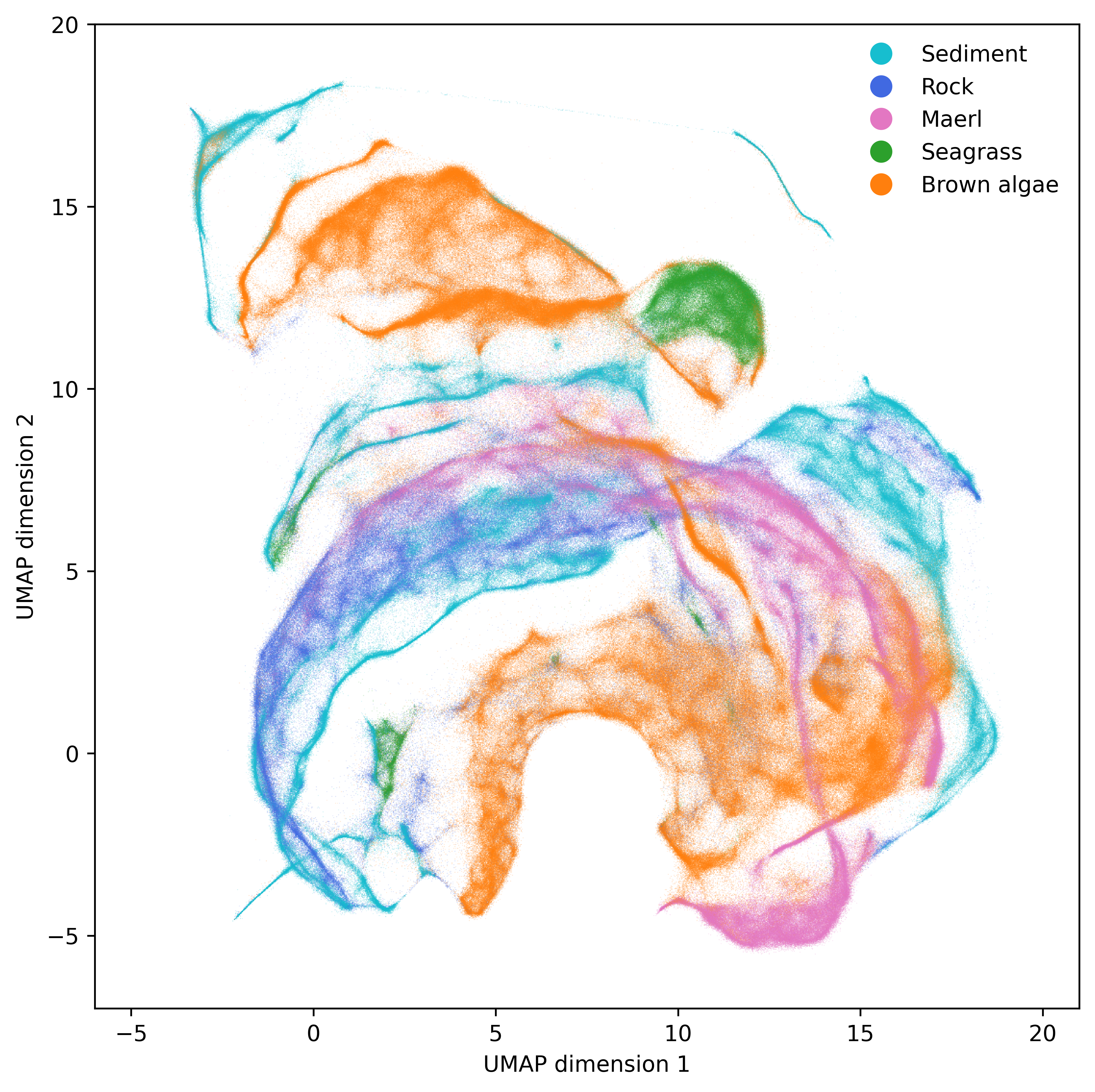}
\caption{Visualization of example spectra from 5 grouped annotation classes, reduced to 2 dimensions via PCA and UMAP.}
\label{fig:umap}
\end{figure}

\section{Technical Validation}\label{sec:technical-validation}
The hyperspectral camera and downwelling irradiance sensor were calibrated by Resonon, and independent calibration and validation was not within the scope of the MASSIMAL project. However, the dataset includes an example image of a "gray card", which is usually used for color balancing in photography, but was used in the project as a reference with known reflectance properties. A short analysis of this image is included here to demonstrate that the values of the hyperspectral image match those of the reference object. The full analysis is available as a Jupyter notebook in \cite{Skjelvareid2026a}.

Figure \ref{fig:reflectance-validation} shows the image, with zoomed-in versions showing the foldable 50×50 cm 18\% reflectance card (Caruba, Beilen, the Netherlands), and plots showing radiance, irradiance and reflectance spectra. The upper plot shows the mean radiance values for the gray card, with the 5th and 95th percentiles representing the range of variation. The percentile band widens above 950 nm, reflecting the low signal-to-noise ratio in this spectral region. The middle plot presents the downwelling irradiance spectrum measured by a spectrometer with a cosine collector. 

The reflectance values in the lower plot closely approximate the expected 18\% level, with a minor underestimation between 400–550 nm, likely due either to a small systematic error in the measurements, or slightly non-flat reflectance properties of the gray card itself.

Small spectral spikes occur near atmospheric absorption features, particularly around 760 nm (Fraunhofer line A), caused by slight wavelength mismatches between radiance and irradiance spectra. These effects can be partly mitigated by smoothing, although reflectance values in a narrow range around 760 nm are best excluded. Above 900 nm, reduced signal-to-noise causes reflectance values to diverge. 

\begin{figure}[!t]
    \centering
    \begin{minipage}[t]{0.46\linewidth}
        \centering
        \includegraphics[width=\linewidth]{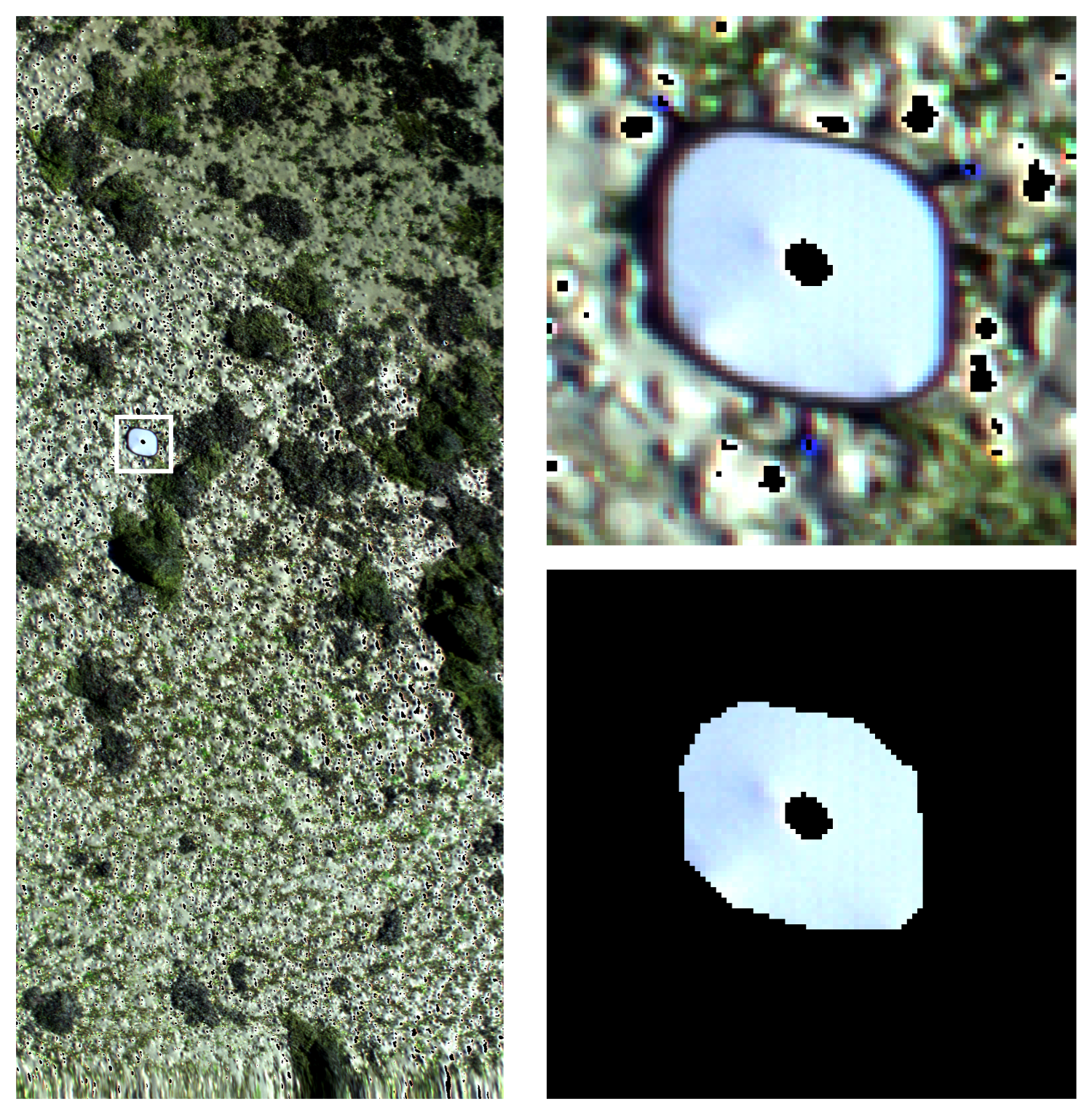}
    \end{minipage}
    \begin{minipage}[t]{0.5\linewidth}
        \centering
        \includegraphics[width=\linewidth]{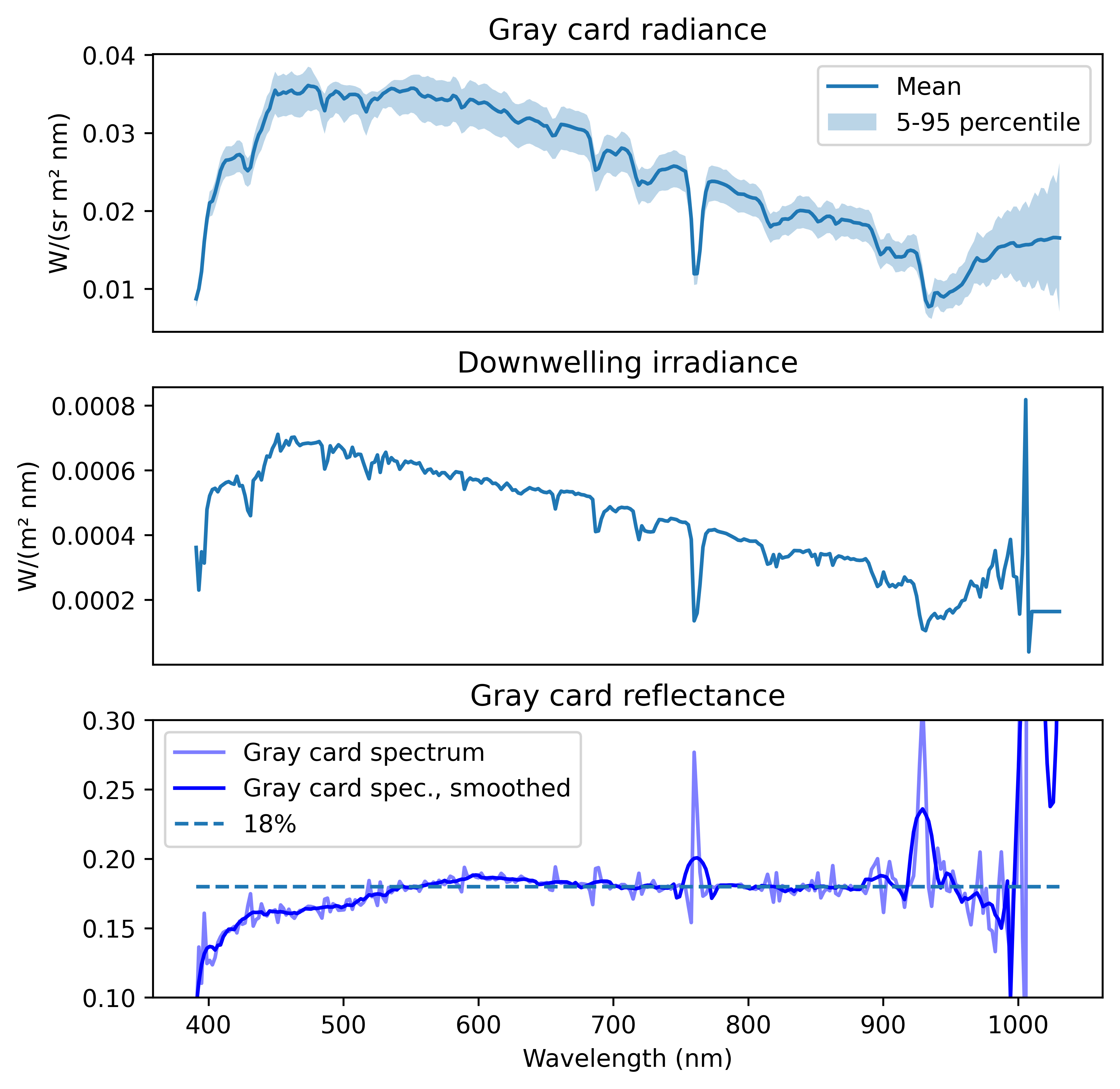}
    \end{minipage}
    \caption{Example image with reference gray card placed on a beach among clusters of macroalgae. Black spots in the image correspond to invalid or saturated pixels in the original radiance data. The image was taken in Juvika, Bodø (file \filename{massimal_bodo_juvika_202104181131-transects-15m_hsi_002}) }
    \label{fig:reflectance-validation}
\end{figure}

\section{Usage Notes}\label{sec:usage-notes}
\subsection{Generating reflectance images} \label{sec:generating-reflectance}
The radiance images published in the dataset, which include measurements of downwelling irradiance where available, can be used to generate reflectance images. The easiest way to do this is via the MassiPipe pipeline, and a notebook demonstrating this is available in \cite{Skjelvareid2026a}. Note, however, that the calculation can be done using any tool that enables reading of ENVI files and manipulating the resulting arrays.

\subsection{Loading annotations} \label{sec:loading-annotations}
Annotations for hyperspectral images are available in two formats; PNG images with annotation masks, and JSON files using the COCO annotation format. A notebook demonstrating how to read both formats and displaying the corresponding annotations is available in \cite{Skjelvareid2026a}.

\subsection{Glint correction} \label{sec:glint-correction}
The MASSIMAL dataset includes multiple examples of images with pronounced sun and sky glint, i.e. reflections of light in the water surface. Glint injects noise in the dataset that is unrelated to the seabed and water column. The MassiPipe pipeline implements a method, described by Hedley et al. \cite{Hedley2005}, to remove glint from hyperspectral images. The method is based on a linear regression model that estimates and removes glint in the visible wavelength range based on the near-infrared light. The model should be trained on example data from images of deep water, i.e., parts of the image where the upwelling light from the water is weak and spatially homogeneous. The \filename{config.seabee.yaml} file that is part of each hyperspectral dataset include image numbers and bounding boxes for training the glint correction model, and glint corrected datasets can thus be generated directly from these, using MassiPipe. 

Figure \ref{fig:glint} shows examples of two images before and after glint correction. Image a) is from a shallow area with macroalgae, seagrass and sand, and was captured on a day with clear skies and some wind. Small waves on the water surface create small areas of glint. Image c) is from a reef in an area where where kelp is harvested commercially, and was captured on a day with overcast skies and calm waters. The downwelling light from the sky is much more diffuse than in image a), which makes the patterns of glint more diffuse.  In image b) and d) the glint is removed, making the details of the seabed much more visible. This is particularly striking in image d), which reveals linear patterns of exposed bedrock from kelp trawling, and shallow kelp forest in the upper-left corner. The glint correction method is not perfect, as exemplified by the image artifacts along the right edge of image d). 

\begin{figure}
\centering
    \includegraphics[width=\linewidth]{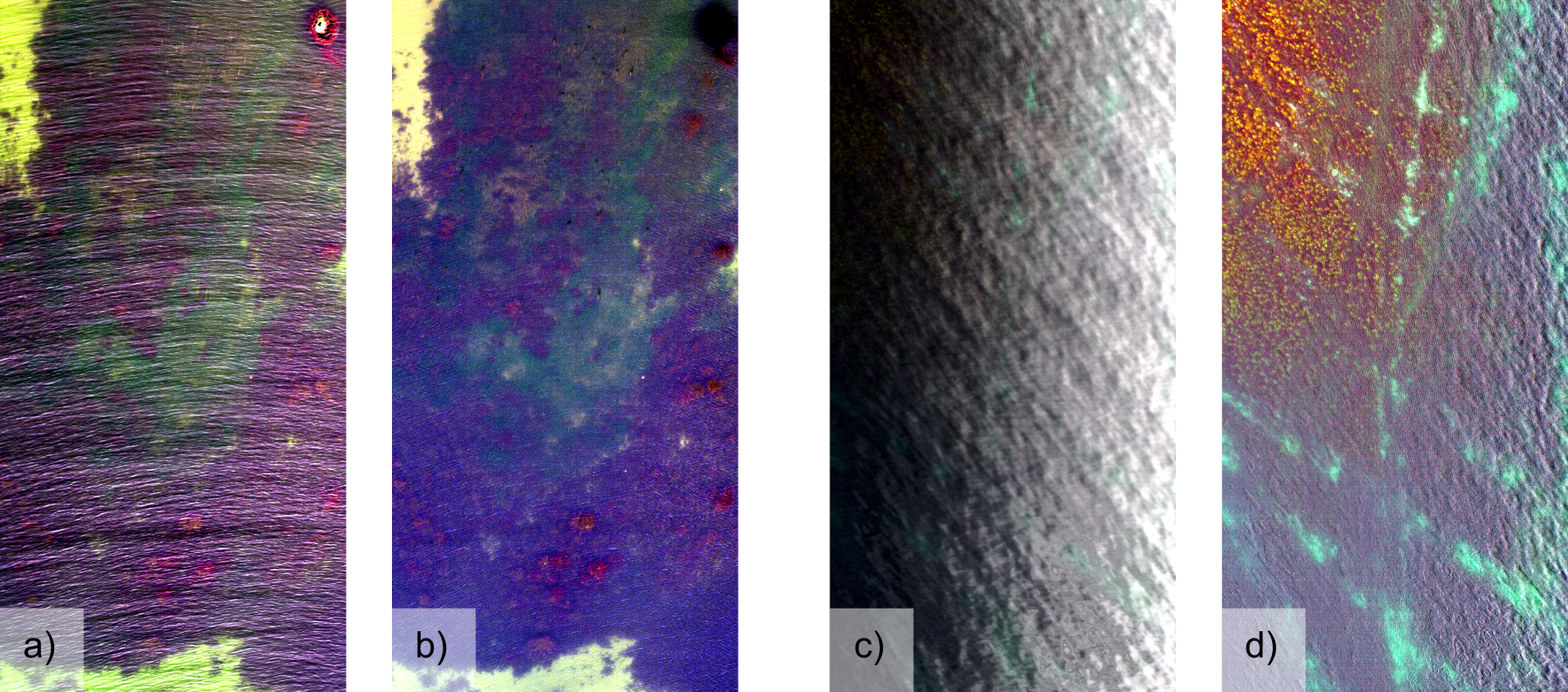}
    \caption{Two examples of images with glint (a,c) and the corresponding images with glint removed (b,d). All images are RGB representations of hyperspectral images, contrast stretched individually per band. Image a) was captured at Larvik, Ølbergholmen (\filename{massimal_larvik_olbergholmen_202308300959-south1_hsi_008})  and image c) at Smøla, Måholmen (\filename{massimal_smola_maholmen_202306211129-2_hsi_009}).}
\label{fig:glint}
\end{figure}

For a more detailed explanation and demonstration of glint correction, which also includes a comparison of spectra before and after correction, see the example notebook in \cite{Skjelvareid2026a}.

\section{Data Availability}\label{sec:data-availability}
The MASSIMAL dataset is publicly available through the NIRD Research Data Archive, under the CC-BY-4.0 license \cite{CreativeCommons2013}. The data can be accessed at \url{https://doi.org/10.11582/2025.00041} ~\cite{Skjelvareid2025}. The total size of the dataset is 1.1 TB, and to allow scientists to access and use the only part of the data at a time, the data has been saved as a set of compressed ZIP files, with each file corresponding to a single dataset from a single modality, as described in the Section \ref{sec:data-records}. Datasets with hyperspectral images are generally the largest, with typical file sizes of 10-20 GB.  

The data can be downloaded via HTTP protocol or the S3 protocol, as described in the dataset repository documentation \cite{NrisDatasetDownload2026}. An online table of contents lists URLs for each zip file, for both protocols, and the HTTP URLs can be clicked for direct download of single files. Using the S3 protocol requires freely available command-line tools such as rclone (Rclone Services Ltd, Guildford, Surrey, England) or AWS CLI (Amazon Web Services, Seattle, Washington, USA). Note that these tools enable downloading of the entire dataset, or parts of the dataset within a subfolder (e.g., all data from one location), without explicitly using the URLs for each individual file.  

All UAV images (hyperspectral, multispectral, and RGB) can also be accessed via the SeaBee Geo-Visualization Portal \cite{SeaBeeGeoPortal2026} (search for "massimal"). The portal provides web map service (WMS) URLs which can be used to display images in GIS applications. Signed-in users can also download data from the portal.

\section{Code Availability}\label{sec:code-availability}

\subsection{MassiPipe}\label{sec:massipipe}
A pipeline for post-processing hyperspectral images was developed as part of the MASSIMAL research project. The pipeline, called \texttt{MassiPipe}, was used to prepare the hyperspectral data published in the present dataset. It can also be used to derive additional image products from the hyperspectral data, such as reflectance images and glint-corrected images. The version of the software used to process the dataset (v0.3.0) is available via Zenodo \cite{MartinSkjelvareid2025a}. MassiPipe is now being developed as a more general-purpose tool for post-processing aerial hyperspectral images, and is available on GitHub (\url{https://github.com/mh-skjelvareid/massipipe}). 

\subsection{VidTransGeoTag}\label{sec:vidtransgeotag}
The ground truth observations in the dataset includes a large number of geotagged images, which were created from two separate streams of data: 1) video from a camera, and 2) a log of geographic positions, measured via a separate, high-resolution GNSS system. By synchronizing the data streams via their time stamps, it is possible to extract images from the video stream and assign geographic positions to them. The extraction was performed via a separate Python tool, named \texttt{vidtransgeotag}, which is available on GitHub (\url{https://github.com/mh-skjelvareid/vidtransgeotag}). The specific version used for preparing the dataset is available via Zenodo \cite{MartinSkjelvareid2025}.

\printbibliography

\newpage

\section{Author Contributions}\label{sec:author-contributions}
(Entered directly in the submission system, not stated here)

\section{Competing Interests}\label{sec:competing-interests}
(Entered directly in the submission system, not stated here)

\section{Acknowledgements}\label{sec:acknowledgements}
(Entered directly in the submission system, not stated here)

\section{Funding}\label{sec:funding}
The MASSIMAL project was funded by the Research Council of Norway (grant number 301317, 8 million NOK) and UiT the Arctic University of Norway (600 000 NOK). Cofunding is acknowledged from the Norwegian Infrastructure for drone-based research, mapping, and monitoring in the coastal zone (SeaBee), funded by the Research Council of Norway (grant number 296478). Processing of hyperspectral images for publication was performed on resources provided by Sigma2 - the National Infrastructure for High-Performance Computing and Data Storage in Norway.

\section{Ethics Statement}\label{sec:ethics-statement}
This study involved the remote sensing of shallow marine habitats using exclusively non-invasive techniques. No direct interaction with or disturbance of marine life or habitats occurred at any stage of data collection or analysis. As such, the research posed no risk to organisms, ecosystems, or the surrounding environment. All methods complied with relevant ethical standards for environmental research, and no permits for biological sampling or habitat manipulation were required.

\end{adjustwidth}

\end{document}